\documentclass[10pt,conference]{IEEEtran}
\IEEEoverridecommandlockouts
\usepackage{booktabs}
\usepackage{threeparttable}
\usepackage{multirow}
\usepackage{xspace}
\usepackage{cite}
\usepackage{amsmath,amssymb,amsfonts}
\usepackage{algorithmic}
\usepackage{graphicx}
\usepackage{textcomp}
\usepackage{xcolor}
\usepackage{url}
\usepackage{booktabs} % For formal tables
\usepackage{tabularx}
\usepackage{graphicx}
\usepackage{caption}
\usepackage{subcaption}
\usepackage{amsmath}
\usepackage{balance} 
\usepackage{amssymb}
\usepackage{xspace}
\usepackage[T1]{fontenc}
\usepackage[scaled=0.81]{beramono}
\usepackage{balance}

\usepackage{enumitem}  
\usepackage{listings}
\usepackage{url}
\usepackage[ruled,lined,linesnumbered,vlined]{algorithm2e}
\usepackage{multirow}
\usepackage{array}
\usepackage{color}
\usepackage{float}
\usepackage[misc]{ifsym}
\newcommand{\myparagraph}[1]{\vspace*{0.14cm}\textbf{#1.}\quad}
\newcommand{\tool}{{Kaya}\xspace}
\newcommand{\etal}{{\emph{et al.}}\xspace}
\newcommand{\eg}{{\emph{e.g.}}\xspace}
\newcommand{\ie}{{\emph{i.e.}}\xspace}
\newcommand{\etc}{{\emph{etc.}}\xspace}

\def\BibTeX{{\rm B\kern-.05em{\sc i\kern-.025em b}\kern-.08em
		T\kern-.1667em\lower.7ex\hbox{E}\kern-.125emX}}
\begin{document}

% \title{Kaya: Effective and Convenient Testing Framework for DApps}
% \title{Kaya: An Effective and Convenient Tool for Comprehending and Testing DApps}
\title{Kaya: A Testing Framework for Blockchain-based Decentralized Applications}

% \author{Zhenhao Wu}
% \affiliation{%
% 	\institution{Peking University}
% 	\city{Beijing}
% 	\country{China}
% }
% \email{zhenhaowu@pku.edu.cn}

% \author{Jiashuo Zhang}
% \affiliation{%
% 	\institution{Peking University}
% 	\city{Beijing}
% 	\country{China}
% }
% \email{zhangjiashuo@pku.edu.cn}

% \author{Jianbo Gao}
% \affiliation{%
% 	\institution{Peking University}
% 	\city{Beijing}
% 	\country{China}
% }
% \email{gaojianbo@pku.edu.cn}

% \author{Yue Li}
% \affiliation{%
% 	\institution{Peking University}
% 	\city{Beijing}
% 	\country{China}
% }
% \email{liyue_cs@pku.edu.cn}

% \author{Qingshan Li}
% \affiliation{%
% 	\institution{Peking University}
% 	\city{Beijing}
% 	\country{China}
% }
% \email{liqs@pku.edu.cn}

% \author{Zhi Guan}
% \affiliation{%
% 	\institution{Peking University}
% 	\city{Beijing}
% 	\country{China}
% }
% \email{guan@pku.edu.cn}

% \author{Zhong Chen}
% \affiliation{%
% 	\institution{Peking University}
% 	\city{Beijing}
% 	\country{China}
% }
% \email{zhongchen@pku.edu.cn}

\author{
	\IEEEauthorblockN{
		Zhenhao Wu\IEEEauthorrefmark{1}, 
		Jiashuo Zhang\IEEEauthorrefmark{1},
		Jianbo Gao\IEEEauthorrefmark{1},
		Yue Li\IEEEauthorrefmark{1}, 
		Qingshan Li\IEEEauthorrefmark{1}, 
		Zhi Guan\IEEEauthorrefmark{1},
		Zhong Chen\IEEEauthorrefmark{1}\IEEEauthorrefmark{3}
	}
	\IEEEauthorblockA{\IEEEauthorrefmark{1}\textit{School of Electronics Engineering and Computer Science}, \textit{Peking University}, Beijing, China} 
	\IEEEauthorblockA{\IEEEauthorrefmark{3}Corresponding Author}
	\{zhenhaowu,zhangjiashuo,gaojianbo,liyue\_cs, liqs,guan,zhongchen\}@pku.edu.cn
	
}

\maketitle
\begin{abstract}
	 In recent years, many decentralized applications based on blockchain (DApp) have been developed.
    % In recent years, blockchain has been widely accepted by academia and industry due to its extraordinary advantages. A lot of decentralized applications based on blockchain (DApp) are also developed to bring these advantages to end-users. 
    However, due to inadequate testing, DApps are easily exposed to serious vulnerabilities. We find three main challenges for DApp testing, \ie the inherent complexity of DApp, inconvenient pre-state setting, and not-so-readable logs. 
    % However, there are some challenges which make DApp testing difficult, \ie the inherent complexity of DApp, inconvenient pre-state setting, and not-so-readable logs. 
    % However, due to some indeed requirements of DApp testing, it is still a challenging task for many test engineers to effectively test DApps. 
    % Firstly, \tool provides an easier way to write and execute test cases. 
% <<<<<<< HEAD
    % Secondly, \tool maintains the stability of the runtime environments, \ie, Blockchain pre-state, to ensure the reliability of testing results. 
    % Thirdly, \tool transforms meaningless addresses into readable variables for easy comprehension.
    % % enables test engineers to write test cases in an easier way with DApp Behavior Description Language (DBDL). Secondly \tool performs test cases automatically. Thirdly, \tool transforms not-so-readable logs into a format that test engineers can comprehend. 
    % With these functions, \tool implements the properties of easy testing, trusted testing and log comprehension. So, test engineers can test DApps more easily with \tool. To fit the various application environments, we provides two ways for test engineers to use \tool, \ie UI and command-line. Our experimental case demonstrates the potential of \tool in helping test engineers to test DApps more easily\footnote{Video's url: https://www.youtube.com/channel/UCrY4b9-WeCPyWDn-58Na4Eg}.
% =======
	In this paper, we propose a testing framework named Kaya to bridge these gaps. Kaya has three main functions.
	Firstly, Kaya proposes DApp behavior description language (DBDL) to make writing test cases easier. Test cases written in DBDL can also be automatically executed by Kaya. 
    Secondly, \tool supports a flexible and convenient way for test engineers to set the blockchain pre-states easily.
    %eliminate undesired changes of Blockchain state to ensures the reliability of testing results. 
    Thirdly, \tool transforms incomprehensible addresses into readable variables for easy comprehension.
    % enables test engineers to write test cases in an easier way with DApp Behavior Description Language (DBDL). Secondly \tool performs test cases automatically. Thirdly, \tool transforms not-so-readable logs into a format that test engineers can comprehend. 
    With these functions, \tool can help test engineers test DApps more easily.
    % With these functions, \tool supports easy testing and comprehensive logs. 
    Besides, to fit the various application environments, we provide two ways for test engineers to use \tool, \ie, UI and command-line. Our experimental case demonstrates the potential of \tool in helping test engineers to test DApps more easily\footnote{Video's url: https://www.youtube.com/watch?v=2OlYV-5deAk}.

\end{abstract}

\section{Introduction}
\label{sec:1.intro}
Since the decentralized cryptocurrency BitCoin was proposed by Satoshi Nakamoto \cite{nakamoto2008bitcoin}, blockchain, smart contracts, and DApps have developed rapidly. % and attracted lots of attention from academia and industry.
%Since blockchain has no centralized control but are maintained according to decentralized consensus models, decentralized applications (DApp) are also very popular recently. 
% At this moment, DApp is also proposed. 
DApp is an application which usually uses a browser as its front-end and uses smart contracts as its back-end. DApp enables end users to interact directly with blockchain, \eg exchanging tokens and playing a blockchain-based game. With this property, DApps have become really popular in market.
% In general, DApp has been widely accepted with the spread of blockchain.
% provides more practical features and better user experience, thus it has a large market.
% This kind of DApp is an application that enables end users to interact directly with blockchain (\eg storing files in blockchain, playing a blockchain-based game). % TODO: 参考文献，看健博
% DApp is an application that enables end users to interact with blockchain directly and conveniently. 
% Different from traditional DApps running on the Peer-to-Peer network, blockchain-based DApps can provide more practical features, better user experience, and safer data storage. Thus, this new kind of DApps is widely accepted with blockchain.
% Different from those traditional DApps running on Peer-to-Peer (P2P) network, like BitTorrent,  DApps are based on blockchain, especially Ethereum.
According to the report \cite{dappcom2019} published by
\textit{dapp.com}, the annual transaction volume of DApps based on blockchain has reached up to 10 billion dollars in 2019.  % TODO: 仅仅表明DApp市场的庞大。

% and Ethereum is the first choice for engineers. 
% TODO: 感觉不是很需要下面的这部分，因为和 testing 无关
% In another word, Ethereum has become the biggest platform of DApps based on blockchain. A reason for this situation is the presence of quasi-Turning-complete virtual machine (EVM) \cite{buterin2013whitepaper} and Remix\footnote{http://remix.ethereum.org/}, they provide a more convenient environment to program smart contracts for DApps based on blockchain.

% Ethereum has become the biggest platform of blockchain-based DApps due to its quasi-Turning-complete virtual machine \cite{buterin2013whitepaper}, which provides a convenient environment for developers to program smart contracts. Now, DApps have been adopted to almost all areas, the total market has reached 3.28 billions of dollors as of January 2019, and Ethereum is the number one choice for developers \cite{dappcom2019Q2}.

% 为了方便开发者们进行快速的 DApp 开发，有许多帮助进行 DApp 开发和部署的工具被提了出来，比如，Remix 和 Truffle, Remix 是
% 为了适应如火如荼的 DApp 开发热潮，市面上也出现了许多对 DApp 开发有利的工具，比如，Remix 和 Truffle。但是，这一类工具主要注重于 Smart Contract 的开发、测试和部署，主要面向开发者。发行一个 powerful 的 DApp, 全面的测试工作也是非常重要的。由于缺乏全面的测试，许多已经发行的 DApp 都暴露除了严重的漏洞。
Before publishing a powerful DApp, comprehensive testing is essential. % for the reliability and security. 
However, due to the difficulty of executing test cases on DApps, many DApps are lack of comprehensive testing and some of them are exposed to serious vulnerabilities, \eg the Parity wallet was attacked in July 2017 which causes the loss of 31M dollars and MyEtherWallet was attacked in April 2018 and lost 17M dollars. 
%DApp has become a new security bottleneck of blockchain. The DAO was attacked by reentrancy vulnerability in June 2016 which leads to the Ethereum's hard fork; the Parity wallet was attacked by Access Control in July 2017; BecToken was attacked due to the integer overflow in April 2018. These events all cause large tokens transferred into malicious accounts and other bad consequences like a hard fork.
% DApp 是 浏览器/区块链 架构的程序，它的逻辑部分通过智能合约来实现，交互接口则通过网络应用的形式来开发。只关注网络应用或者只关注智能合约的测试方法都难以直接应用在 DApp 的测试中。

DApp testing involves not only users' behaviors but also the logic of smart contracts \cite{gao2019towards}.
% DApp is a Browser/blockchain architecture program whose back-end logic is implemented through associating smart contracts,%(引用高健博的论文), 
%However, the front-end interfaces are developed through the form of web applications. 
Hence, recent popular smart contract testing tools \eg Remix\footnote{http://remix.ethereum.org/} and Truffle\footnote{https://www.trufflesuite.com/}, are not suitable for DApp testing. These tools focus on the development, testing, and deployment of smart contracts, \ie, these tools are trying to help smart contract developers rather than DApp test engineers.
Through the investigation of DApp ecology, we find three main challenges and list them as follows. 

\myparagraph{Challenge 1: Inherent complexity of DApp} In practice, it is necessary for test engineers to test all relevant aspects of DApp at the same time, including front-end events, back-end logic, and some deeper restrictions. This means testing DApps is more complex than testing other applications. However, there is a lack of effective tools to reduce the complexity.

% TODO: 需要更好的描述
% <<<<<<< HEAD
% \myparagraph{Challenge 2: Untrusted testing results} Blockchain pre-state is the runtime environment of the back-end smart contract of DApp. It consists of all previous information recorded in Blockchain. Due to this property, the results of previous test cases also participate in constructing the Blockchain pre-state, \i.e., the Blockchain pre-state of the previous test case is different from the  the latter test case. The difference of Blockchain pre-state could make the results differ from what they should be. Unfortunately, we have not found a testing tool that can eliminate such differences.

% % Blockchain has a popular property that records all information forever. However this property may influence the reliability of DApp testing results, especially when executing similar test cases with the same Blockchain account. During the process of testing, the results of previous test case will be recorded to participate in the execution of the latter test case, \ie, the Blockchain state of the previous test case is different from the Blockchain state of the latter test case. This situation could make the results differ from what they should be.
% =======

% \myparagraph{Challenge 2: Requirements for convenient and flexible pre-state setting} 
\myparagraph{Challenge 2: Inconvenient pre-state setting} Blockchain is stateful. Different pre-states may lead to different execution resulst. During testing, there are some test cases should be executed in certain pre-states, \ie, integer overflow. Existing method to simulate an certain pre-state is redeploying the smart contract and setting state variables' value seperately using their incomprehensible addresses. It is costly and inconvenient. How to conveniently and flexibly change these pre-states is still an unsolved problem.
% Smart contracts on blockchain are stateful. For the same set of test cases, different execution sequences or different pre-states may lead to different results. However, certain bugs may only be triggered in certain pre-states. To find those bugs, the pre-state of smart contract are expected to be set to an certain state before test engineers executing the test case. 
% However, for tests engineers, the only way to do this is redeploying the smart contract and setting state varibles' values separately using there incomprehensible addresses. It is costly and inconvenient. 
% How to convenietly and flexibly change these pre-states is still an unsolved problem.
%has a popular property that records all information forever. However this property may influence the reliability of DApp testing results, especially when executing similar test cases with the same blockchain account. During the process of testing, the results of previous test case will be recorded to participate in the execution of the latter test case, \ie, the blockchain state of the previous test case is different from the blockchain state of the latter test case. This situation could make the results differ from what they should be.
% >>>>>>> 9c716c365a4db3eeb8f3b5176052dc738326497c

% 挑战二：难以理解的日志输出。
\myparagraph{Challenge 3: Requirements for comprehensive logs} In traditional programs, logs are readable and easy to understand. % With these logs, we can locate most bugs easily. 
However, In DApps, logs produced by smart contract virtual machine (SCVM)\footnote{The most common way to implement smart contracts is the virtual machine approach. There are various virtual machines, \eg, EVM, Neo VM, Move VM, \etc. We refer to these virtual machines as smart contract virtual machines, SCVM} are not as readable as traditional logs. The variables in SCVM logs are displayed as addresses rather than readable names, and the operations taken by SCVM are shown in a format similar to assembly code. Both of them make SCVM logs difficult to understand.

In this paper, we propose a testing framework called \tool\footnote{\tool is an item in game DOTA2, it has two main buffs, loss reduction and damage amplification. Our testing framework can also help test engineers to reduce the difficulties and improve the efficiency} to help test engineers test DApps more easily. \tool implements three core functions to solve above three challenges. % These three core functions are listed.
\begin{enumerate}
    %自动执行测试用例，既可以进行前端操作，也可以运行后端智能合约，前端的操作行为产生的数据可以直接在运行智能合约的时候使用 Simplify test case writting and execution
    % TODO: 需要精细修改
% <<<<<<< HEAD
%     \item Provide an easier way to write and execute test cases with DBDL. In writing, DBDL can easily add front-end events, back-end smart contracts, and Blockchain states. In executing, \tool will execute all test cases automatically, including simulating front-end events, running smart contracts, and setting Blockchain state.
%     %允许测试工程师以一种更简单的方式来撰写测试用例，且测试用例中可以涵盖对前端和后端的测试行为
%     \item Maintain the stability of Blockchain Pre-state. 
%     %Eliminate undesired changes of Blockchain state. 
%     Before executing a test case, \tool re-constructs the Blockchain pre-state according to relevant parameters. This operation can eliminate the impact of the previous test case, and make the latter test case be executed under the same Blockchain pre-state as the previous test case.
% =======
    \item Provide an easy way for testing by writing test cases with DBDL and executing test cases automatically. Using DBDL, test engineers can easily pack those complex front-end events and back-end smart contracts into simple test cases. \tool will execute these test cases automatically, \ie simulating front-end events, setting blockchain pre-states and running smart contracts.
    %允许测试工程师以一种更简单的方式来撰写测试用例，且测试用例中可以涵盖对前端和后端的测试行为
    \item Provide a flexible and convenient way for test engineers to set the blockchain pre-state as they need. Before executing test cases, \tool will construct the blockchain pre-state according to the settings of test engineer. 
% >>>>>>> 9c716c365a4db3eeb8f3b5176052dc738326497c
    %which is acceptable for \tool to perform DApp testing.
    % Enable test engineers to write feature-rich test cases in an easier way with DApp behavior description language (DBDL).
    %支持对已定义数据的修改，以检验DApp对恶意攻击的抵抗能力
    % \item Support to change pre-defined data to test the resistance to malicious attacks.
    %可以对区块链的输出进行分析，让执行结果易于测试工程师进行理解。
    \item Generate readable analysis report for easy comprehension. \tool can calculate the correspondence between addresses and variables and transform not-so-readable logs produced by SCVM into a form which is much easier to understand for test engineers.
    % \item \label{enum:logic} Turn meaningless addresses into meaningful expression, reduce the difficulty of understanding the program. % Analyze used variables to help test engineers quickly and better understand the primary logic of smart contract. With this function, test engineers can 
    % \item \label{enum:tool} Generate effective test cases in simple DApp behavior description language (DBDL). These cases can contain defined user behaviors on this DApp.
    % \item \label{enum:logic-tool}Perform test cases automatically to identify DApp logic and help test engineers to understand the function and logic of smart contracts better.
\end{enumerate}

% To improve the testing technology on DApps, there are two main challenges:
% \begin{enumerate}
%     \item How to generate test cases effectively even by normal testers.
%     \item How to automatically perform test cases in spite of the existing of incomprehensible addresses.
% \end{enumerate}

% We proposed a testing framework called \tool\footnote{\tool is an item in game DOTA2, it has two main buffs, loss reduction and damage amplification. Our testing framework can also help testers to reduce the workload and improve the efficiency}\footnote{https://github.com/Ghands/Kaya}, which not only allow testers to design their own test cases with our defined language, DApp behavior description language (DBDL), but also can perform these test cases automatically. 

The first function simplifies the work of writing and executing test cases, which solves the Challenge 1. The second function makes test cases can be executed in a certain state to solve the Challenge 2. The third function overcomes Challenge 3 by helping test engineers to analyze the results of DApp testing like normal software testing. In this paper, we conduct experimental cases to evaluate the effect of \tool. These cases show that \tool is indeed effective in DApp testing.
% The first function makes the execution of DApp-related test cases become automatic, the second function reduces the difficulty of writing test cases and the third function helps test engineers analyze DApp logs with traditional methodology. In other words, the union of the first two functions solves Challenge 1 while the third function overcomes Challenge 2. To evaluate the effect of \tool, we conduct experimental cases, the results show that \tool is indeed effective in DApp testing.

% In addition to providing practical functions for comprehending and testing on DApp, \tool is also convenient in usage. It supports multiple methods of deployment and remote access, test engineers can start their testing in a browser without installing even one software.
% Besides, a raw test engineer can carry out testing following the steps we designed while an experienced test engineer can use the command tool to directly perform his test cases written by DBDL.
% With the help of \tool, test engineers can test a DApp like traditional software, they can divert attention from understanding blockchain and smart contract related knowledge to generating more comprehensive and effective test cases.
% With the help of \tool, test engineers can focus on testing more functions of a DApp to improve the correctness, reliability and security.

% testers can pay more attention to generating better test cases to cover more functions of a DApp, to improve the correctness, reliability, coverage, and security like on traditional software.

\section{Overview}
\label{sec:2.overview}
\begin{figure}[htb]
    \centering
    \includegraphics[width=0.5\textwidth]{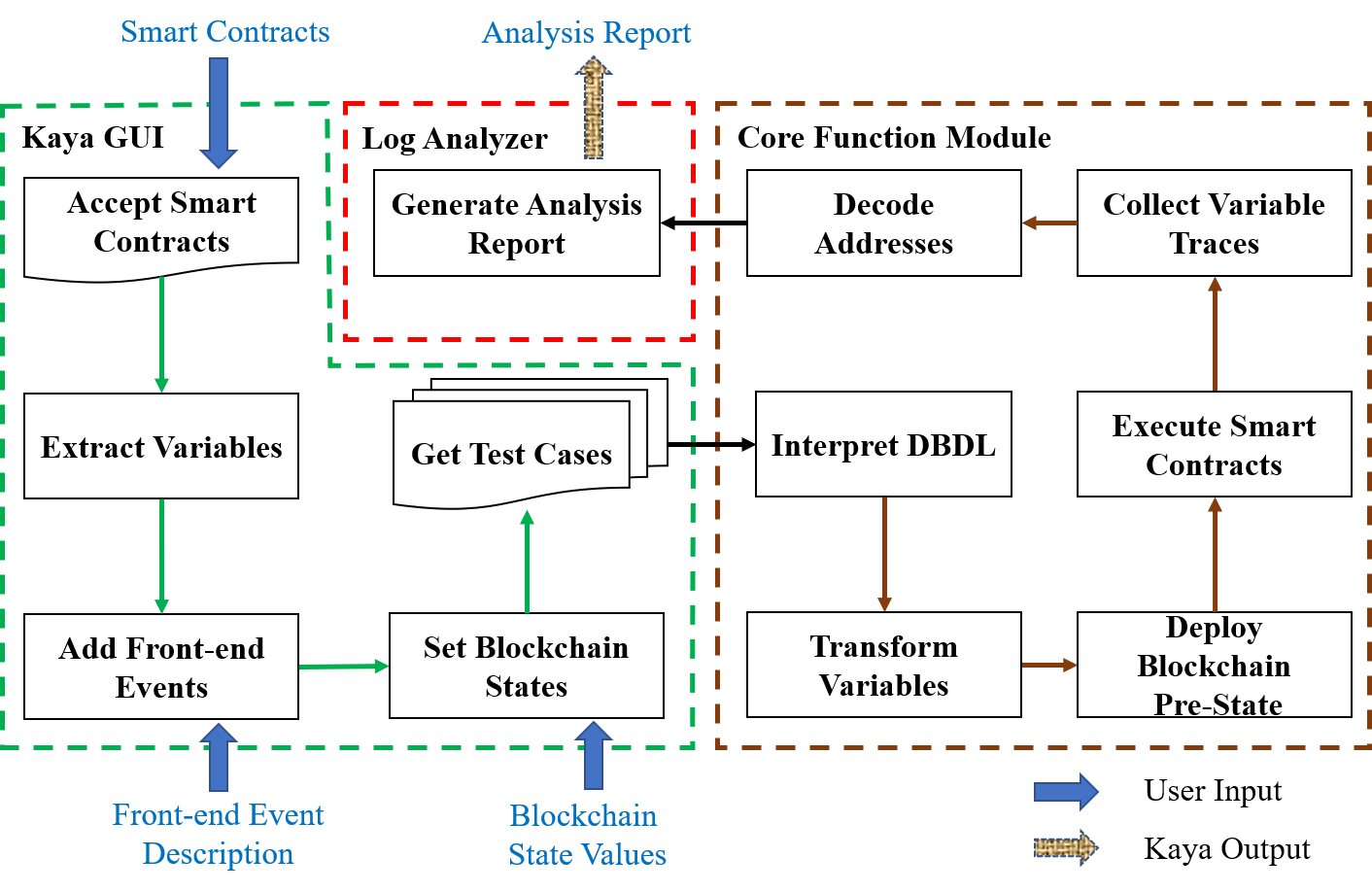}
    \caption{The framework and workflow of \tool}
    \label{fig:framework}
\end{figure}

In this section, we will introduce  the  framework and workflow of  \tool.
\tool is composed of three modules, \ie  \tool Graphical User Interface (KGUI), Core Function Module (CFM), and Log Analyzer (LA). Each of these three modules plays a unique role in Kaya. KGUI helps test engineers to write test cases with DBDL easily and quickly. CFM executes test cases and transforms variables into addresses. LA generates readable analysis reports for test engineers.
% Except the  Core Function Module, the others have a certain sequence. 
The relationship between these three modules in \tool is illustrated as Fig.~\ref{fig:framework}. 

\subsection{\tool Graphical User Interface}
KGUI accepts inputs from test engineers and outputs test cases written in DBDL to CFM. DBDL is a domain specific language created for describing test cases in \tool, the details of DBDL are introduced in Section~\ref{sec:3.dbdl}. The inputs in KGUI should be the smart contracts used by the target DApp, front-end events description, and blockchain pre-state parameters. 

% These inputs will be aggregated into test cases written in DBDL. 
% \myparagraph{DApp Interfaces} \label{mod:1-1} In DApp Interfaces, test engineers can specify the DApp to be tested.
KGUI follows the following steps to work.
\begin{enumerate}
    \item \myparagraph{Accept Smart Contracts} At the beginning of this module, test engineers need to input the source codes of smart contracts used in the target DApp.
    % \item \myparagraph{Smart Contract Analysis} Then, the source codes of smart contracts are analyzed to ensure their availability.
    \item \myparagraph{Extract Variables} Then, all useful variables contained in the source codes of smart contracts are extracted. These variables will become the parameters when setting  pre-state. % for test engineers can simulate different runtime states of smart contracts.
    % TODO: 
    % Next, all useful variables are extracted for test engineers to view. 
    % Test engineers can change their values freely to simulate different states of Smart Contracts.
    \item \myparagraph{Add Front-end Events} Next, test engineers can add front-end events to describe the usage of the target DApp. These events should be written in DBDL.
    \item \myparagraph{Set Pre-states} Test engineers can also set some parameters to construct the blockchain pre-states. The constructed pre-states will be the runtime environment of those test cases. Settable parameters include the variables extracted from smart contracts and the attributes of used blockchain accounts, \eg the balance.
    \item \myparagraph{Get Test Cases} Finally, front-end events and blockchain pre-state parameters are packed into test cases written in DBDL. These test cases are transmitted to CFM for further execution.
    % Finally, front-end events are performed to obtain some additional information about the target DApp. The obtained information and Blockchain state parameters both participate in generating test cases written in DBDL. 
\end{enumerate}

 % A DApp may involve multiple smart contracts, and test engineers should input the source code of the smart contract to be tested. For a public DApp, the smart contracts are visible in BlockChain, such as Ethereum; for a publishing DApp, test engineers can acquire the smart contracts from developers.
% Then, Test Case User Interface analyzes the source code with its function, Smart Contract Analysis. Next, Variable Interpreter extracts all available variables for test engineers to view and set. Test engineers can also add front-end behaviors to help \tool extract some information about DApp usage, or hack extracted variables with changing values to simulate malicious attacks in Web Events/Variables Setting. Finally, Test Case User Interface transforms all settings into the test case with DBDL format.

% \myparagraph{Web Events/Variables Setting} \label{mod:1-3} To make the testing more comprehensive and effective, Test Case Helper accepts two types of input, one is front-end behaviors, another is the changed initial value of variables. Front-end behaviors can help \tool to extract some information about DApp usage while changing the initial value of variables can simulate the malware attack to test the hidden restriction of a DApp.

% The information of variables used in \ref{mod:1-3} are analyzed from the input smart contract by functional part \ref{mod:2-1} in Core Function Module. The final output of Test Case Helper are test cases written in DBDL. % TODO: 说明DBDL的章节

\subsection{Core Function Module}

CFM is the key module in \tool. It takes the responsibilities of executing test cases automatically, including collecting useful information, 
% extracting useful information contained in test cases and smart contracts, 
transforming blockchain pre-state parameters into (address, value) pairs, executing test cases, converting the output of SCVM to a readable format.
% (address,value) pairs for execution.
% DBDL Module is a key module of \tool, all test cases are written with it, 
% With DSL Module, we can transform test cases into Pre-states
% , which contain the key information of each test case, and can be used by EVM. In addition, DSL Module also plays an important role in interpreting the output of EVM
The workflow of this module is shown as follows.

\begin{enumerate}
    \item \myparagraph{Interpret DBDL} This step accepts test cases written in DBDL and translates them into the data structure which is convenient for the next step to handle. During interpretation, front-end events are executed by \tool to obtain necessary information.
    \item \myparagraph{Transform Variables} In this step, all blockchain pre-state parameters are transformed into (address, value) pairs in SCVM according the address calculation rules. Without this step, SCVM cannot recognize these parameters.
    \item \myparagraph{Deploy Blockchain Pre-State} Blockchain pre-state are the runtime environment of test cases. Before running a test case, CFM will deploy the blockchain pre-state according to the setted parameters. 
    % Parameters with address format will be used to construct the pre-state in Blockchain. Pre-state is the aggregation of all state elements, \eg, the balance in user account, the goods user owned, the restrictions predefined in smart contracts, \etc. Only with the fixed pre-state, all test cases can be executed correctly.
    \item \myparagraph{Execute Smart Contracts} Run smart contracts in SCVM with the pre-state. 
    \item \myparagraph{Collect Variable Traces} Collect all variable traces produced by SCVM for further analysis.
    \item \myparagraph{Decode Addresses} This step transforms addresses into the origin names of variables. After this step, logs are readable for test engineers.
\end{enumerate}

\subsection{Log Analyzer}

LA analyzes the outputs generated by CFM. After analysis, a report is produced to show changes of each variable. With this report, test engineers can easily check the program logics implemented by the target DApp.

% Without Address Decoder, test engineers have to analyze the meaningless addresses themselves, which is really difficult for most test engineers, especially it can be an NP-hard problem to find variable names on some occasions.

% Log Analyzer can analyze logs generated by Contract Executor with the help of Address Decoder. 
% DSL Decoder is able to transform variables from the format of addresses to the format of names. 
% Without Address Decoder, testers have to understand the meanings of addresses, which is really difficult for most testers and it can even become an NP-hard problom to find the variables name in some occasions.

% Log Analyzer will produce a report to show the change of each variable. Combined with the user behaviors, testers can easily understand the testing results, judge related functions in a DAPP have been implemented right or not.

% \section{Tool Usage}
\section{DApp Behavior Description Language}
\label{sec:3.dbdl}
% General introduce

\begin{figure}[htb]
    \centering
    \includegraphics[width=0.45\textwidth]{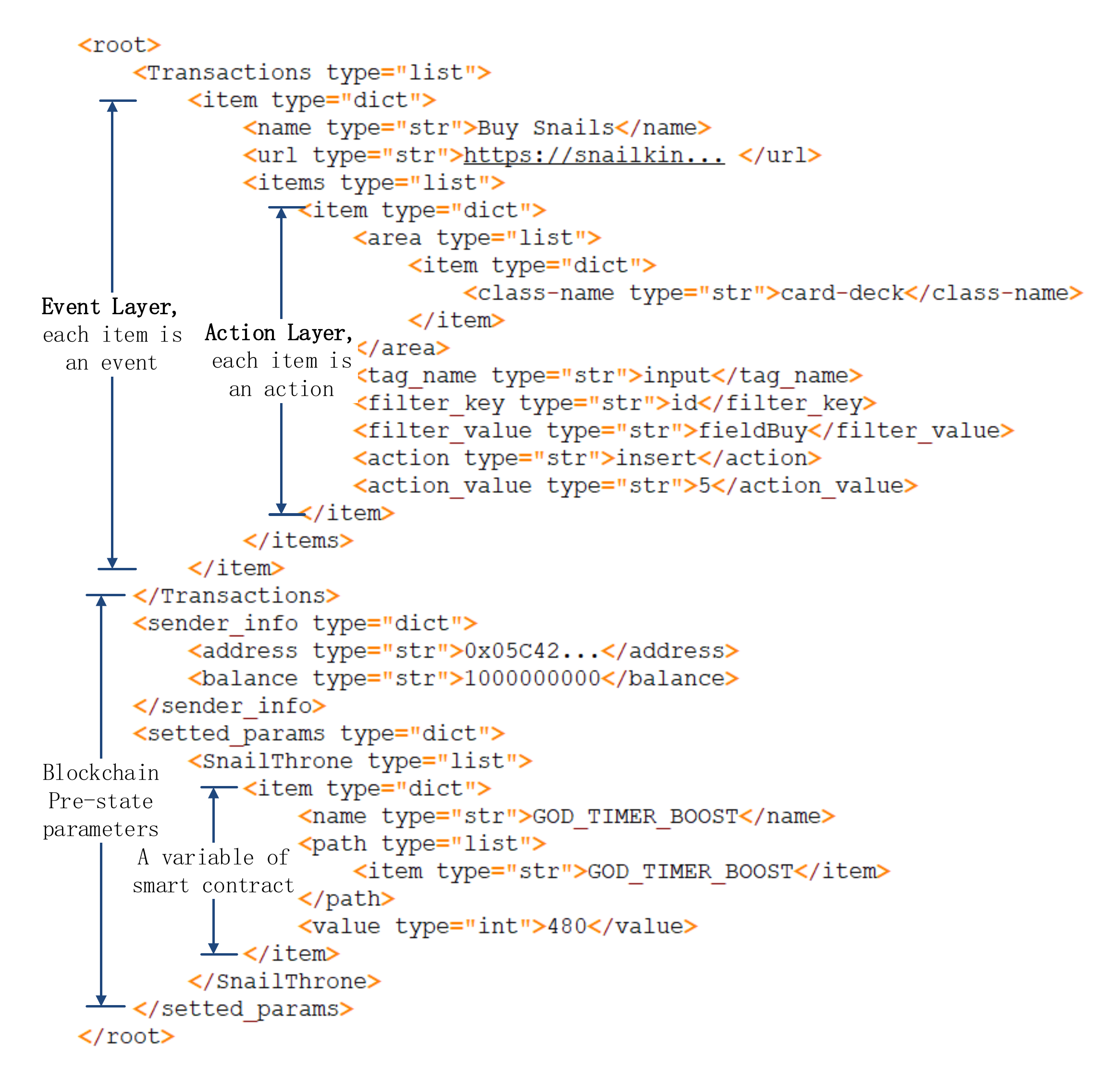}
    \caption{A test case written in DBDL}
    \label{fig:dbdl-ex}
\end{figure}

DBDL is created for writing test cases conveniently in Kaya. 
% It is flexible enough to 
With DBDL, test engineers can easily add front-end events and pre-state parameters to simulate various scenarios of DApp usage. An example of a test case written in DBDL is shown in Fig.~\ref{fig:dbdl-ex}.

In DBDL, a test case can be divided into three main fields, \ie  \texttt{Transactions} with front-end events, \texttt{sender\_info} with account attributes, and \texttt{setted\_params} with variables of smart contracts. \texttt{sender\_info} and \texttt{setted\_params} form the blockchain pre-state parameters. %These three fields correspond to front-end behaviors, user account status, and variables with changed initial values.

The content in field \texttt{Transactions} is consist of two layers. The outer layer is Event Layer, which represents front-end events. Each front-end event has a self-defined event name (\texttt{name}), an access address (\texttt{url}), and an Action Layer. An event may be composed of several actions, so there may be multiple actions in Action Layer. Each action consists of the target area (\texttt{area}), target tag name (\texttt{tag\_name}), filter (\texttt{filter\_key} and \texttt{filter\_value}), and the taken action (\texttt{action\_key} and \texttt{action\_value}) to find the target element and perform the defined action. If there are multiple events and actions, \tool will perform them in order, \ie from beginning to end.

Field \texttt{sender\_info} records the pre-state of the blockchain account. Attributes of this account can directly be used as the tag in this field.

Field \texttt{setted\_params} records the variables of smart contracts. These variables are grouped by the name of the smart contract they belong to. Each tag \texttt{item} means a variable, including a self-defined name, access path, and the value user want to set (the default value is zero).

DBDL allows to define the values of variables in smart contracts. We require that all tags except \texttt{root} should declare their attributes by \texttt{type} to ensure the normal execution of test cases.

\section{Case Study}
\label{sec:4.case}
In this section, we will provide two cases to show the features of \tool. The tested DApp is SnailThrone\footnote{https://snailking.github.io/snailthrone/game}, which is a King of the Hill game on Etheruem. Related source codes that integrate EVM into \tool have been released\footnote{https://github.com/Ghands/Kaya}. % and we integrate EVM with its address calculation methods in these two cases.

It is worth noting that Kaya is a testing framework. It is not limited to a specific blockchain platform. To apply it to other blockchain platforms, developers need to implement the key modules of Kaya according to the SCVM used in the target blockchain platform.

\subsection{Case with User Interface}

\begin{figure}[htb]
    \centering
    \includegraphics[width=0.4\textwidth]{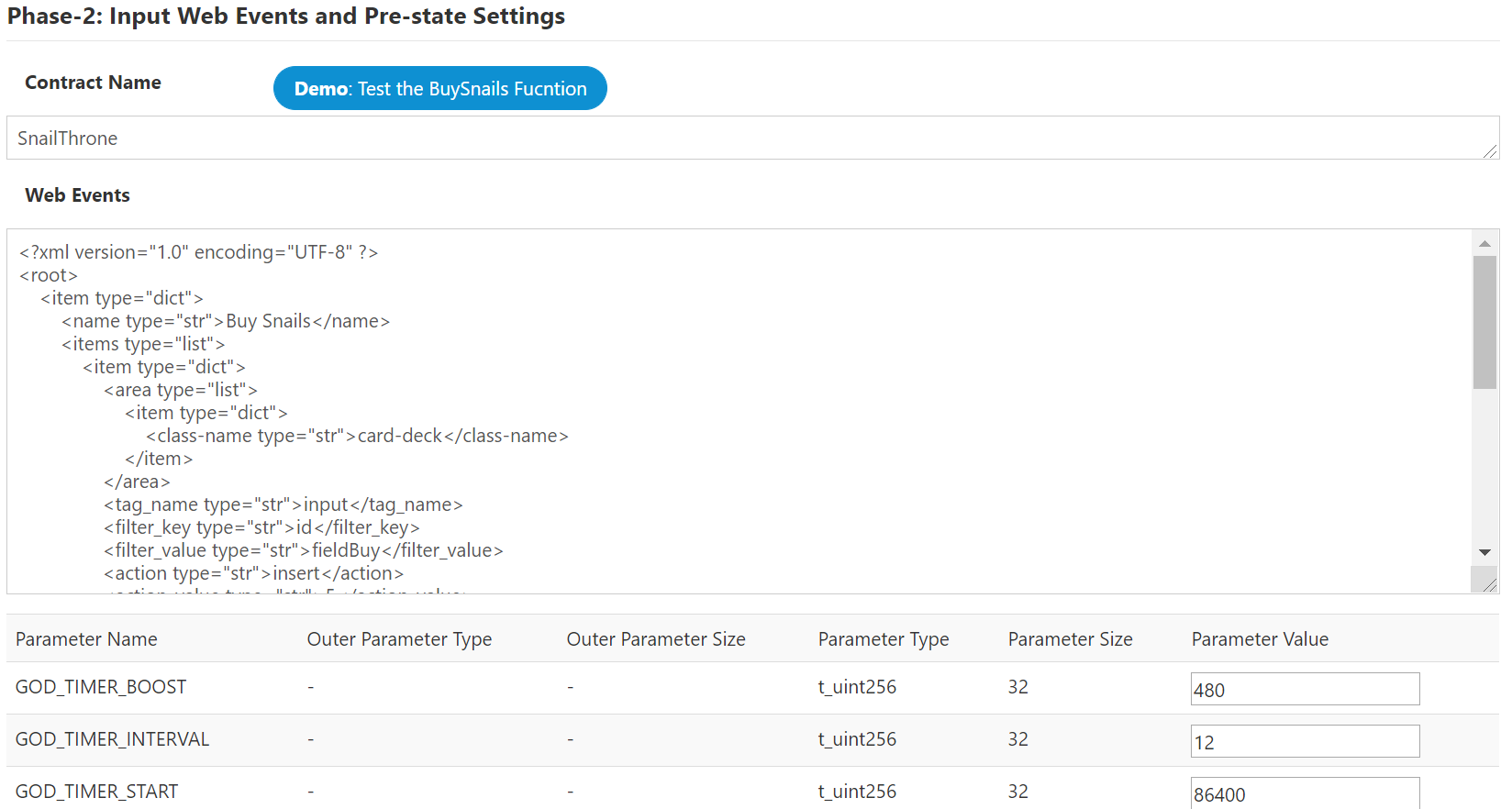}
    \caption{The key step to generate test cases}
    \label{fig:case-gui}
\end{figure}

KGUI of \tool guides test engineers to test a DApp step by step. At first, test engineers should input the name and source code of a smart contract. Then \tool analyzes the smart contract to provide a list of useful variables. Test engineers can add front-end events to simulate user behaviors or set blockchain pre-state parameters to obtain expected blockchain pre-state. This step is shown in Fig.~\ref{fig:case-gui}.

\begin{figure}[htb]
    \centering
    \includegraphics[width=0.45\textwidth]{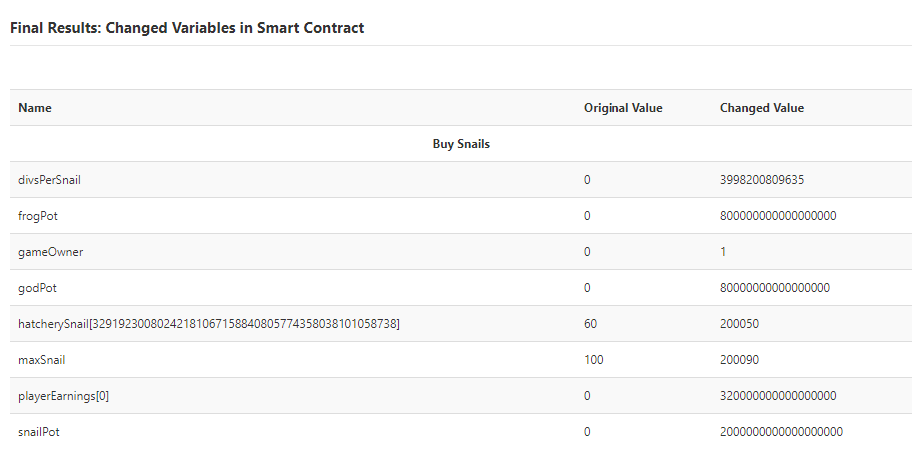}
    \caption{The analysis report generated by \tool}
    \label{fig:analysis-report}
\end{figure}

Next, \tool extracts useful information from front-end events and constructs pre-states to execute test cases.
% Next, \tool will start to obtain related information about web events, which is used to generate EVM Pre-states with these variables changed initial value. 
Finally, \tool outputs the analysis report like Fig.~\ref{fig:analysis-report}.

When the whole process finishes, test engineers can identify the features of DApp from all output variables.
% the changes between input variables and output variables. 
For example, from Fig.~\ref{fig:analysis-report}, we can know the earning of a player (playerEarnings[0]) may have a positive correlation with the number of snails (hatcherySnail[$3\dots38$]) he has. 

\subsection{Case with Command-line Tool}

\begin{figure}[htb]
    \centering
    \includegraphics[width=0.45\textwidth]{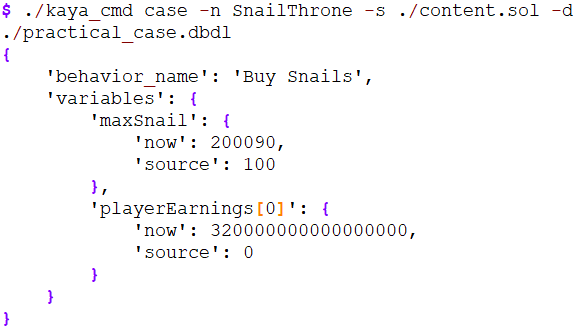}
    \caption{Run Kaya command tool}
    \label{fig:kaya-cmd}
\end{figure}

Except for the user interface, experienced test engineers can also use the command-line tool to perform test cases with \tool. This command-line tool called \texttt{kaya\_cmd}, which can analyze the variables contained in a smart contract and perform test cases.
% with front-end events and Blockchain pre-state parameters. 
To improve the efficiency, \texttt{kaya\_cmd} supplies a more practical function: test engineers can directly input the source code of smart contracts, front-end events, and Blockchain pre-state parameters together to skip the steps in KGUI. Fig.~\ref{fig:kaya-cmd} shows a case with this command-line tool, the output is as simple and clear as it showed in the user interface. % With the result, test engineers can 

% For experienced test engineers, the normal tester interface is not efficient enough. \tool also provides a command tool called \texttt{kaya\_cmd}, it can also analyze the variables contained in a smart contract and perform test cases with tester defined initial values. The usage and final outputs are shown in. Parameter \texttt{var} and \texttt{case} correspond to analyzing variables and performing test cases. The final output is shown with the Format JSON, the key \texttt{source} means the initial value when this test case starts, the key \texttt{now} means the final value when this test case ends.

\section{Related Work}
\label{sec:5.related}
% To the max knowledge we have, there is no previous work focusing on the state-independent DApp testing, nor use Interface Description Languages for DApp testing. Our work applies Interface Description Languages to DApp Testing and supports state-independent testing.
% \subsection{Web Application  Testing}

% Our work touches several related ares including program testing and interface description language.

\myparagraph{Program Testing} Our work is related to DApp testing, web application testing and smart contract analysis. In DApp Testing, Gao \etc \cite{gao2019towards} proposed an automated testing technique called Sungari for DApps. It achieves significant optimization compared to the random testing approach. For Web Applications, Artzi \etc \cite{artzi2011framework} proposed a framework for feedback-directed automated test generation for JavaScript and implemented a tool called Artemis. Billes \etc \cite{billes2017systematic} presents the first fully automated technique called Simian for multi-client web applications. Smart contract analysis mainly focuses on discovering vulnerabilities. Luu \etal \cite{luu2016making} develop Oyente to detect some types of smart contract bugs with a symbolic execution based technique. Zeus \cite{kalra2018zeus} is a tool to analyze smart contract with format verification. In addition, fuzzing \cite{jiang2018contractfuzzer} is also introduced into smart contract analysis.

%\cite{billes2017systematic}.  Most of those studies focused on improving the coverage and efficiency \cite{jensen2013automated} \cite{sen2013jalangi} \cite{saxena2010symbolic}. 
% \subsection{Interface Description Languages for Testing}

\myparagraph{Interface Description Language} Jensen \etc \cite{jensen2013server} proposed an IDL for the automated testing of JavaScript web applications and incorporated it into the testing tool Artemis. Mahmud \etc \cite{mahmud2010lowering} proposed an easy-to-understand scripting language for test engineers to create test scripts that lower the barriers to the website testing. Rajan \etc \cite{rajan2009weave} created an extensive user interface in their tool to specify requirements for web applications easily.

\section{Conclusion}
\label{sec:6.conclusion}
In this paper, we propose a DApp testing framework called \tool, which aims to help test engineers test DApps more easily. %\tool provides an easier way to write and execute test cases and decrease testing difficulty. 
\tool provides an easy way for testing by writing test cases with DBDL and executing test cases automatically, provide a flexible and convenient way for test engineers to set the blockchain pre-state as they need, generate readable analysis report for easy comprehension.
% complex test cases automatically, provides a convenient and flexible way to write powerful test cases with DBDL, and displays meaningful information instead of the original incomprehensible logs.
To the best of our knowledge, \tool is the first framework which helps test engineers to test DApps in an easier way. We hope the correctness, reliability, and security of DApps can be enhanced by using \tool. In the future, we will continue to improve the ecology of DApp by improving the efficiency and coverage of DApp testing.

% \begin{acks}
% Acknowledgment
% \end{acks}

\bibliographystyle{IEEEtran}
\bibliography{dsl}

\end{document}